\documentstyle[twocolumn,aps,floats]{revtex}

\begin{document}

\draft \tolerance = 10000

\setcounter{topnumber}{1}
\renewcommand{\topfraction}{0.9}
\renewcommand{\textfraction}{0.1}
\renewcommand{\floatpagefraction}{0.9}

%Fixing abstract in twocolumn mode
\twocolumn[\hsize\textwidth\columnwidth\hsize\csname
@twocolumnfalse\endcsname

\title{Does Special Relativity Have Limits of Applicability
in the Domain of Very Large Energies?}
\author{L.Ya.Kobelev  \\ Department of  Physics, Urals State University \\
 Av. Lenina, 51, Ekaterinburg 620083, Russia \\  E-mail: leonid.kobelev@usu.ru}
\maketitle

\begin{abstract}
We have shown in the paper that for time with fractional
dimensions (multifractal time theory) there are small domain of
velocities $v$ near $v=c$ where SR must be replaced by fractal
theory of almost inertial system that doe's not contains an
infinity and permits moving with arbitrary velocities.
\end{abstract}

 \vspace{1cm}

%Fixing abstract in twocolumn mode
]

\section{Introduction}

The special relativity theory (SR) is one of the physics theories that are
compose the base of the modern physics, it has well experimental
foundation in the large area of the reached velocities and energies, is
the working theory of a modern physics and widely use in a science and
technique. Nevertheless, as any physical theory created by men, SR has a
boundaries of applicability (the inertial systems of reference). As far as
we know a problem of energy boundaries SR for any moving body was not
analyzed in detail. At the same time the tends of energy of a moving body
when it's velocity  reaches the value of speed of light to infinity calls
doubts in applicability  SR in this area of energies, as the occurrence of
infinity in the physical theories always testifies about their interior
deficiencies. The purpose of the paper is presenting  the theory of almost
inertial systems in the space with multifractal time (I believe that time
and space of our world may have multifractal characteristics) in which the
motions of bodies with arbitrary velocities are possible and there are no
energy infinity. At the same time if the energies reached by a body, as a
result of it's motion, are smaller than $E_{0}10^{3}sec^{1/2}$ $t^{-1/2}$
 (${t}$-time of acceleration of a particle up to such energies, $E_{0}$- a
rest energy) all results of the theory coincide with results of SR  and
thus do not contradict experiment. The differences between our theory and
SR appear only if energy of moving body exceeds the energy
$E_{0}$$10^{3}$$sec^{1/2} t^{-1/2}$. The theory is based on using, but as
a good approach, of a principle of the constancy speed of light, an
invariance of modified Galilean and Lorentz transformation laws. This
theory is not a generalization of SR, because any SR generalizations in
the domain of validity SR (inertial systems) are absurd. This theory
describes relative movements only in the almost inertial systems, and thus
does not contradict SR, thou coincide with it for case of inertial
systems. For constructing such theory it is necessary to refuse from the
rigorous realization of the SR postulates: a homogeneous of space and
time, the constancy of speed of light, the Galilean relativity principle.
In an inhomogeneous space and time, if the inhomogeneous are small, the
motion of bodies will be almost inertial, and velocity of light is almost
stationary value. This paper contains the example of the theory based on
the time and the space with fractional dimensions (FD)\cite{kob1} . This
theory  use for description of time and space characteristics  the ideas
of fractal geometry. The values FD are a little bit distinguished from the
integer dimensions (the theory of multifractal time and space is given in
 \cite{kob1}. In fractal theory the motions of particles with arbitrary
velocities are permissible, the speed of light is almost independent from
the velocity of lights sources. For example, on  the surface of Earth the
differences of value of speed of light under change moving direction $v$
by $-v$ consist $\sim 2v/c 10^{-6}t$. The theory almost coincides with SR,
for velocities  which  are lesser than speed of light but does not
includes singularities at $v=c$.

\section{Multifractal time}

Following \cite{kob1}, we will consider both time and space as the initial
real material fields existing in the world and generating all other
physical fields. Assume that every of them consists of a continuous, but
not differentiable bounded set of small elements (elementary intervals,
further treated as "points"). Consider the set of small time elements
$S_{t}$. Let time be defined on multifractal subsets of such elements,
defined on certain measure carrier $\mathcal{R}^{n}$. Each element of
these subsets (or "points") is characterized by the  fractional (fractal)
dimension (FD) $d_{t}({\mathbf r}(t),t)$ and for different elements FD are
different. In this case the classical mathematical calculus or fractional
(say, Riemann - Liouville) calculus \cite{sam} can not be applied to
describe a small changes of a continuous function of physical values
$f(t)$, defined on time subsets $S_{t}$, because the fractional exponent
depends on the coordinates and time. Therefore, we have to introduce
integral functionals (both left-sided and right-sided) which are suitable
to describe the dynamics of functions defined on multifractal sets (see
\cite{kob1}). Actually, these functionals are simple and natural
generalization of the Riemann-Liouville fractional derivatives and
integrals:
\begin{equation} \label{1}
D_{+,t}^{d}f(t)=\left( \frac{d}{dt}\right)^{n}\int_{a}^{t}
\frac{f(t^{\prime})dt^{\prime}}{\Gamma
(n-d(t^{\prime}))(t-t^{\prime})^{d(t^{\prime})-n+1}}
\end{equation}
\begin{equation} \label{2}
D_{-,t}^{d}f(t)=(-1)^{n}\left( \frac{d}{dt}\right)
^{n}\int_{t}^{b}\frac{f(t^{\prime})dt^{\prime}}{\Gamma
(n-d(t^{\prime}))(t^{\prime}-t)^{d(t^{\prime})-n+1}}
\end{equation}
where $\Gamma(x)$ is Euler's gamma function, and $a$ and $b$ are some
constants from $[0,\infty)$. In these definitions, as usually, $n=\{d\}+1$
, where $\{d\}$ is the integer part of $d$ if $d\geq 0$ (i.e. $n-1\le
d<n$) and $n=0$ for $d<0$. If $d=const$, the generalized fractional
derivatives (GFD) (\ref{1})-(\ref{2}) coincide with the Riemann -
Liouville fractional derivatives ($d\geq 0$) or fractional integrals
($d<0$). When $d=n+\varepsilon (t),\, \varepsilon (t)\rightarrow 0$, GFD
can be represented by means of integer derivatives and integrals. For
$n=1$, that is, $d=1+\varepsilon$, $\left| \varepsilon \right| <<1$ it is
possible to obtain:
\begin{equation} \label{3}
D_{+,t}^{1+\varepsilon }f(t)\approx \frac{\partial}{\partial t}
f(t)+a\frac{\partial}{\partial t}\left[\varepsilon (r(t),t)f(t)\right]
\end{equation}
where $a$ is constant and defined by the choice of the rules of
regularization of integrals (\ref{1})-(\ref{2}) (for more detailed see
\cite{kob1}). The selection of the rule of regularization that gives a
real additives for usual derivative in (\ref{3}) yield $a=0.5$ for $d<1$
and $a=1.077$ for $d>1$ \cite{kob1}. The functions under integral sign in
(\ref{1})-(\ref{2}) we consider as the generalized functions defined on
the set of the finite functions \cite{gel}. The notions of GFD, similar to
(\ref{1})-(\ref{2}), can also be defined and for the space variables
${\mathbf r}$.

The definitions of GFD (\ref{2})-(\ref{2}) are formal until the
connections between fractal dimensions of time $d_{t}({\mathbf r}(t),t)$
and certain characteristics of physical fields (say, potentials $\Phi
_{i}({\mathbf r}(t),t),\,i=1,2,..)$ or densities of Lagrangians $L_{i}$)
are determined. Following \cite{kob1}, we define this connection by the
relation
\begin{equation} \label{4}
d_{t}({\mathbf r}(t),t)=1+\sum_{i}\beta_{i}L_{i}(\Phi_{i} ({\mathbf
r}(t),t))
\end{equation}
where $L_{i}$ are densities of energy of physical fields, $\beta_{i}$ are
dimensional constants with physical dimension of $[L_{i}]^{-1}$ (it is
worth to choose $\beta _{i}^{\prime}$ in the form $\beta _{i}^{\prime
}=a^{-1}\beta _{i}$ for the sake of independence from regularization
constant). The definition of time as the system of subsets and definition
the FD $d$ (see  \ref{4}) connects the value of fractional (fractal)
dimension $d_{t}(r(t),t)$  with each time instant $t$. The latter depends
both on time $t$ and coordinates ${\mathbf r}$. If $d_{t}=1$ (an absence
of physical fields) the set of time has topological dimension equal to
unity. The multifractal model of time allows, as will be shown below, to
consider the divergence of energy of masses moving with speed of light in
the SR theory as the result of the requirement of rigorous validity of the
laws pointed out in the beginning of this paper in the presence of
physical fields (in our theory there are approximate fulfillment of these
laws).

\section{The principle of the speed of light invariance}

Because of the non-uniformity of time in our multifractal model, the speed
of light, just as in the general relativity theory, depends on potentials
of physical fields that define the fractal dimensions of time
$d_{t}({\mathbf r}(t),t)$ (see (\ref{4})). If fractal dimension
$d_{t}({\mathbf r}(t),t)$ is close enough to unity
($d_{t}(r(t),t)=1+\varepsilon, \, \left| \varepsilon \right|<<1$), the
difference of the speed of light in moving (with velocity $v$ along the
$x$ axis) and fixed frame of reference will be small. In the systems that
move with respect to each other with almost constant velocity (stationary
velocities do not exist in the mathematical theory based on the
definitions of GFD (\ref{1}) - (\ref{2})) the speed of light can not be
taken as a fundamental constant. In the multifractal time theory the
principle of the speed of light invariance can be considered only as
approximate. But if $\varepsilon$ is small, it allows to consider a
nonlinear coordinates transformations from the fixed frame to the moving
frame (replacing the Galilean transformations  in non-uniform time and
space), as close to linear (weakly nonlinear) transformations and, thus,
makes it possible to preserve the conservation laws, and all the
invariant's of the Minkowski space, as the approximate laws. Then the way
of reasoning and argumentation accepted in SR theory (see for example,
(\cite{mat})) can also remains valid. Designating the coordinates in the
moving and fixed frames of reference through $x^{\prime}$ and $x$,
accordingly, we write down
\begin{eqnarray}  \label{5} \nonumber
x^{\prime}&=&\alpha (t,x)[x-v(x,t)t(x(t),t] \\
x&=&\alpha^{\prime}(t,x)[x^{\prime}+
v^{\prime}(x^{\prime}(t^{\prime}),t^{\prime}),
\,t^{\prime}(x^{\prime}(t^{\prime}),t^{\prime})
\end{eqnarray}
In (\ref{5}) $\alpha \neq \alpha^{\prime}$ and the velocities $v^{\prime}$
and $v$ (as well as $t$ and $t^{\prime}$) are not equal (it follows from
the inhomogeneous of multifractal time). Place clocks in origins of both
the frame of references and let the light signal be emitted in the moment,
when the origins of the fixed and moving frames coincide in space and time
at the instant $t{{}^1}=t=0$ and in points $x^{\prime} = x = 0$. The
propagation of light in moving and fixed frames of reference is then
determined by equations
\begin{equation} \label{6}
x^{\prime}=c^{\prime}t^{\prime}\,\,\,x=ct
\end{equation}
These equations characterize the propagation of light in both of the
frames of reference at every moment. Due to the time inhomogeneous
$c^{\prime} \neq c$, but since $\left| \varepsilon <<1 \right|$ the
difference between velocities of light in the two frames of reference will
be small. For this case we can neglect by the differences between
$\alpha^{\prime}$ and $\alpha $ and, for different frames of reference
write the expressions for velocities of light (using (\ref{3}) to define
velocity (denote $f(t)=x,\, dx/dt=c_{0}$)). Thus we obtain
\begin{equation} \label{7}
c=D_{+,t}^{1+\varepsilon}x=c_{0}(1-\varepsilon)- \frac{d\varepsilon}{dt}x
\end{equation}
\begin{equation} \label{8}
c^{\prime}=D_{+,t}^{1+\varepsilon^{\prime}}x^{\prime}=
c_{0}(1-\varepsilon)- \frac{d\varepsilon}{dt}x^{\prime}
\end{equation}
\begin{equation} \label{9}
c_{1}= c_{0}(1-\varepsilon)- \frac{d\varepsilon}{dt}x^{\prime}
\end{equation}
\begin{equation} \label{10}
c_{1}^{\prime}= c_{0}(1-\varepsilon)- \frac{d\varepsilon}{dt}x
\end{equation}
The equalities (\ref{9}) and (\ref{10}) appear in our model of
multifractal time as the result of the statement, that in this model all
the frames of reference are absolute frames of reference (because of the
real nature of the time field) and the speed of light depends on the state
of frames: if the frame of reference is a moving or a fixed one, if the
object under consideration in this frame moves or not. This dependence
disappears only when $\varepsilon=0$. Before substitution the relations
(\ref{5}) in the equalities (\ref{7}) - (\ref{10}) (with
$\alpha^{\prime}\approx\alpha$) it is necessary to find out how
$d\varepsilon/dt$ depends on $\alpha$. Using for this purpose equation
(\ref{4}) we obtain:
\begin{equation} \label{11}
\frac{d\varepsilon}{dt}=\frac{d\varepsilon}{d{\mathbf r}} {\mathbf v}
\approx -\sum_{i}\beta_{i}({\mathbf F}_{i} {\mathbf v} +\frac{\partial
L_{i}}{\partial t})
\end{equation}
where ${\mathbf F }_{i} = \frac{dL_{i}} {d{\mathbf r}}$. As the forces for
moving frames of reference are proportional to $\alpha$ we get (for the
case when there is no explicit dependence of $L_{i}$ on time)
\begin{equation}   \label{12}
\frac{d\varepsilon}{dt}\approx -\sum_{i}\beta_{i}{\mathbf F}_{0i} {\mathbf
v} \alpha
\end{equation}

where $F_{0i}$ are the corresponding forces at zero velocity. Multiplying
(\ref{7}) - (\ref{10}) on the corresponding times
$t,\,t^{\prime},\,t_{1},\,t_{1}^{\prime}$ yields the following expressions
\begin{equation} \label{13}
c^{\prime}t^{\prime}=c_{0}t\left[1+\frac{v\sum_{i}\beta_{i}F_{0i}}
{c_{0}}\alpha^{2}ct(1-\frac{v}{c})\right]
\end{equation}
\begin{equation} \label{14}
ct=c_{0}t^{\prime}\left[1+\frac{v\sum_{i}\beta_{i}F_{0i}}
{c_{0}}\alpha^{2}ct(1-\frac{v}{c})\right]
\end{equation}
\begin{equation} \label{15}
c_{1}^{\prime}t_{1}^{\prime}=c_{0}t_{1}\left[1-\frac{v\sum_{i}\beta_{i}F_{0i}}
{c_{0}}\alpha^{2}ct(1-\frac{v}{c})\right]
\end{equation}
\begin{equation} \label{16}
c_{1}t_{1}=c_{0}t_{1}^{\prime}\left[1-\frac{v\sum_{i}\beta_{i}F_{0i}}
{c_{0}}\alpha^{2}ct(1-\frac{v}{c})\right]
\end{equation}
Since in our model the motion and frames of reference are absolute, the
times $t_{1}$ and $t_{1}^{\prime}$ correspond to the cases, when the
moving and fixed frames of reference exchange their roles - the moving one
becomes fixed and vice versa. These times coincide only when
$\varepsilon=0$. The times in square brackets, as well as the velocities,
are taken to equal, because the terms containing them are small as
compared to unity. The principle of invariance of the velocity of light
for transition between the moving and fixed frames of reference in
multifractal time model is approximate (though quite natural, because the
frames of reference are absolute frames of reference). Taking into account
(\ref{5}), the relations (\ref{13}) - (\ref{16}) take the forms
\begin{equation} \label{17}
c^{\prime}t^{\prime}=c\alpha t(1-\frac{v}{c}),\,\,\,\,
c_{1}^{\prime}t_{1}^{\prime}=c\alpha t_{1}(1-\frac{v}{c})
\end{equation}
\begin{equation} \label{18}
ct=c\alpha t^{\prime}(1+\frac{v}{c}),\,\,\,\, c_{1}t_{1}=c\alpha
t^{\prime}(1+\frac{v}{c})
\end{equation}
Once again we note, that the four equations for
$c_{1}^{\prime}t_{1}^{\prime}$ and $c_{1}t_{1}$, instead of the two
equations in special relativity, appear as the consequences of the
absolute character of the motion and frames of reference in the model of
multifractal time. In the right-hand side of (\ref{17}) - (\ref{18}) the
dependence of velocity of light on fractal dimensions of time is not taken
into account (just as in the equations (\ref{13}) - (\ref{16})). Actually,
this dependence leads to pretty unwieldy expressions. But if we retain
only the terms that depend on $\beta=\sqrt{|1-v^{2}/c^{2}|}$ or $a_{0}$
and neglect non-essential terms containing the products $\beta
\alpha_{0}$, utilizing (\ref{13}) - (\ref{16}) after the multiplication of
the four equalities (\ref{17}) - (\ref{18}), we receive the following
equation for $\alpha$ (it satisfies to all four equations):
\begin{equation} \label{19}
4a_{0}^{4}\beta^{4}\alpha^{8}-4a_{0}^{2}\alpha^{4}+1=
\beta^{4}\alpha^{4}+4a_{0}^{4}\beta^{4}\alpha^{8}
\end{equation}
where
\begin{equation} \label{20}
\beta=\sqrt{\left|1-\frac{v^{2}}{c^{2}} \right|}
\end{equation}
\begin{equation} \label{21}
a_{0}=\sum_{i}\beta_{i}F_{0i}\frac{v}{c}ct
\end{equation}
From (\ref{19}) follows
\begin{equation} \label{22}
\alpha_{1}\equiv\beta_{*}^{-1}= \frac{1}{\sqrt[4]{\beta^{4}+4a_{0}^2}}
\end{equation}
The solutions $\alpha_{2,3,4}$ are given by
$\alpha_{2}=-\alpha_{1},\,\alpha_{3,4}=\pm i\alpha$. Applicability of
above obtained results is restricted by requirement $|\varepsilon| \ll 1$

\section{Lorentz transformations and transformations of length \\
and time in multifractal time model}

The Lorentz transformations, as well as transformations of coordinate
frames of reference, in the multifractal model of time are nonlinear due
to the dependence of the fractional dimensions of time $d_{t}({\mathbf
r},t)$ on coordinates and time. Since the nonlinear corrections to Lorentz
transformation rules are very small for $\varepsilon \ll 1$, we shall take
into account only the corrections that eliminate the singularity at the
velocity $v=c$. It yields in the replacement of the factor $\beta^{-1}$ in
Lorentz transformations by the modified factor $\alpha=1/\beta^{*}$ given
by (\ref{22}). The Lorentz transformation rules (for the motion along the
$x$ axis) take the form
\begin{equation}     \label{23}
x^{\prime}=\frac{1}{\beta^{*}}(x-vt),\,\,\,
t^{\prime}\frac{1}{\beta^{*}}(t-x\frac{v}{c^{2}})
\end{equation}
In the equations (\ref{22}) and (\ref{23}) the velocities $v$ and $c$
weakly depend on $x$ and $t$  and their contribution to the singular terms
are small. Hence, we can neglect by this dependence. The transformations
from fixed system to moving system are almost orthogonal (for $\varepsilon
\ll 1$ ), and the squares of almost four-dimensional the energy-momentum
vectors of Minkowski space vary under the coordinates transformations very
slightly (i.e. they are almost invariant). Then it is possible to neglect
the correction terms of order about $O(\varepsilon,\dot{\varepsilon})$,
which, for not equal to infinity variables, are very small too. From
(\ref{22}) - (\ref{23}) the possibility of arbitrary velocity motion of
bodies with nonzero rest mass follows. With the corrections of to order
$O(\varepsilon,\dot{\varepsilon})$ in nonsingular terms being neglected,
the momentum and energy of a body with a nonzero rest mass in the frame of
reference moving along the $x$ axis ($E_{0}=m_{0}c^{2})$ equal to
\begin{equation} \label{24}
p=\frac{1}{\beta^{*}}m_{0}v=\frac{m_{0}v}{\sqrt[4]{\beta^{4}+4a_{0}^{2}}},\,\,\,
E=E_{0}\sqrt{\frac{v^{2}c^{-2}}{\sqrt{\beta^{4}+4a_{0}^{2}}}+1}
\end{equation}
The energy of such a body reaches its maximal value at $v=c$ and is equal
then $E_{v=c}\approx E_{0}/\sqrt{2\alpha_{0}}$. When $v \rightarrow\infty$
the energy is finite an tends to $E_{0}\sqrt{2}$. For $v \le c$ the total
energy of a body is represented by the expression
\begin{equation} \label{25}
E\cong\frac{E_{0}}{\sqrt[4]{\beta^{4}+4 a_{0}^{2}}}=mc^{2},\,\,\,
m=\frac{m_{0}}{\beta^
{*}}
\end{equation}
For $v \ge c$, total energy, defined by (\ref{24}), is given by
\begin{eqnarray}  \label{26}
 m = \beta^{*-1} m_0 \sqrt {\text{1} +\beta^{*2}  +\sqrt {\beta^{*4}
- 4a_0^2 }}
\end{eqnarray}
If we  take into account only the gravitational field of Earth (here, as
in (\cite{log}),the  gravitational field is a real field) and neglect by
the influences of all the other fields), the parameter $a_{0}(t)$ can be
estimated as $a_{0}=r_{0}R^{-3}x_{E}c t$, where $r_{0}$ is the
gravitational radius of Earth, $r$ is the distance from the Earth's
surface to its center ($\varepsilon=0.5\beta_{g}\Phi_{g},\,
\beta_{g}=2c^{-2},\,x_{E}\sim r_{0},\,v=c$). For energy maximum  we get
$E_{max}\sim E_{0}\cdot10^{3}t^{-0.5}sec^{-0.5}$. If we take into account
only  the constant electric field  with electric strange $E$ then
parameter $a_{0}$ has the value:$ a_{0}= \frac{eE}{Mc^{2}}ct$ where
$Mc^{2}$ is the rest energy of charges originated the strange $E$ and $t$
is the time of acceleration of particle.
 Contraction of lengths and the retardation of time
in moving frames of reference in the model of multifractal time are also
have several peculiarities. Let $l$ and $t$ be the length and time
interval in a fixed frame of reference. In a moving frame
\begin{equation} \label{27}
l^{\prime}=\beta^{\ast} l,\,\,\, t^{\prime}=\beta^{\ast}t
\end{equation}
Thus, there exist the maximal contraction of length when the body's
velocity equals the speed of light, but length is not equal to zero. With
the further increasing of velocity (if it is possible to fulfill some
requirements for a motion in this region with constant velocity without
radiating), the length of a body begins to grow and at infinitely large
velocity is also infinite. The retardation of time, from the point of view
of the observer in the fixed system (maximal retardation equals to
$t^{\prime}=t\sqrt{2a_{0}}$) is replaced, with the further increase of
velocity over the speed of light, by acceleration of a flow of time ($t
\to 0$ when $v \to \infty$).

The rule for velocities transformation keeps its form, but $\beta$ is
replaced by $\beta^{*}$
\begin{equation} \label{28}
u_{x}=\frac{u_{x}^{\prime}+v}{1+\frac{u_{x}^{\prime} v}{c^{2}}},\,
u_{y}=\frac{u_{y}^{\prime}\beta^{\ast}}{1+\frac{u_{y}^{\prime}
v}{c^{2}}},\,
u_{z}=\frac{u_{z}^{\prime}\beta^{\ast}}{1+\frac{u_{z}^{\prime} v}{c^{2}}}
\end{equation}
Since there is no law that prohibits velocities greater than that of
light, the velocities in (\ref{28}) can also exceed the speed of light.
The electrodynamics of moving media in the model of multifractal time can
be obtained, in most cases, by the substitution $\beta\to\beta^{*}$.
\section{Newton equation for relativistic particle}
The Newton relativistic  equation for particle with velocities $v\leq c$
has form
\begin{equation}\label{29}
\frac{\partial}{\partial t}(m{\bf v})=\frac{\partial}{\partial t}\left(
\frac{m_{0}{\bf v}}{\sqrt[4]{(1-\frac{v_{2}}{c_{2}})^{2}+ 4a_{0}^{2}}}
\right) =e {\bf E}
\end{equation}
For $v\geq c$ the mass $m$ in equation (\ref{29}) is determined by
(\ref{26}). If ${\bf E}= const.$ (\ref{29}) gives for $v=c$ possibility to
find (neglecting by radiation of charge) minimum for the time $t_{0}$ that
is  necessary for receiving by particles the velocity equal $c$
\begin{equation}\label{30}
 \frac{m_{0}c}{\sqrt{\frac{2eE ct_{0}}{Mc^{2}}}} =eEt_{0}
\end{equation}
or
\begin{equation}\label{31}
  t_{0} = \sqrt[3]{\frac{Mc^{2}}{2E_{0}}}\quad\frac{E_{0}}{eEc}
\end{equation}
 The  $t_{0}$ defined by (\ref{31}) gives only an order of the value $t$
 that is necessary for receiving by particle velocity $v=c$ . The
  maximum energy at $v=c$ may be written now as (we introduce the value
 $\alpha<1$  for describing the  radiation energy losses )
\begin{equation}\label{32}
  E_{max}= E_{0} \sqrt[3]{\frac{Mc^{2}}{2E_{0}\alpha}}
\end{equation}
The values of $\alpha$ and electric strange $E$ are determined by
construction of accelerators and conditions of their work regimes.

 \section{Conclusions}

The theory of relative motions in almost inertial systems based on the
multifractal time theory \cite{kob1} and constructed in this paper gives
in the new describing for characteristics (energy, momentum, mass and so
on) of moving bodies. The main results are:a) the possibility of moving
with arbitrary velocities without appearance of infinitum energy  and
imaginary mass; b) existence of maximum energy if $v=c$; c) possibility of
experimental verification the main results of the theory. This theory
describes open systems (theory of open systems see in { \cite{klim}). The
theory coincides with SR after transition to inertial systems (if neglect
by the fractional dimensions of time) or almost coincides (the differences
are non-essential) for velocities $v<c$. The movement of bodies with
velocities that  exceed the speed of light is accompanied by a series of
physical effect's which can be found experimentally (these effects was
considered in the separate papers (\cite{kob1})  in more details). It is
necessary for verification of the theory to receive the particle's with
energy $E\sim \frac{ E_{0}}{\sqrt{\frac{2eE}{Mc^{2}} ct_{0}}}$ where
$t_{0}$ defined by (\ref{31}). If take into account the  radiation losses
of energy, it will be enough to receive at the  intervals of  time
acceleration of the particle about one second the energies of order $E\sim
E_{0}10^{3}$.

\end{document}